\documentclass[3p,times,procedia]{elsarticle}
\usepackage{nupha_ecrc}

\volume{00}

\firstpage{1}

\journalname{Nuclear Physics A}

\runauth{Iu. Karpenko et al.}

\jid{nupha}

\jnltitlelogo{Nuclear Physics A}

\usepackage{amssymb}

\usepackage[figuresright]{rotating}

\newcommand{\sNN}{s_\mathrm{NN}}

\begin{document}

\begin{frontmatter}



\dochead{}

\title{Effects of EoS in viscous hydro+cascade model for the RHIC Beam Energy Scan}


\author[INFN,BITP,FIAS]{Iu.~Karpenko}
\author[FIAS,ITP]{M.~Bleicher}
\author[FIAS,ITP,WR]{P.~Huovinen}
\author[FIAS,ITP,GSI]{H.~Petersen}
\address[INFN]{INFN - Sezione di Firenze, Via G. Sansone 1, I-50019 Sesto Fiorentino (Firenze), Italy}
\address[BITP]{Bogolyubov Institute for Theoretical Physics, 14-b, Metrolohichna st., 03680 Kiev, Ukraine}
\address[FIAS]{Frankfurt Institute for Advanced Studies, Ruth-Moufang-Stra{\ss}e 1, D-60438 Frankfurt am Main, Germany}
\address[ITP]{Institute for Theoretical Physics, Johann Wolfgang Goethe Universit\"at, Max-von-Laue-Str.~1, D-60438 Frankfurt am Main, Germany}
\address[WR]{Institute of Theoretical Physics, University of Wroclaw, pl.~Maxa Borna 9, PL-50204 Wroclaw, Poland}
\address[GSI]{GSI Helmholtzzentrum f\"ur Schwerionenforschung GmbH, Planckstr. 1,  D-64291 Darmstadt, Germany}

\begin{abstract}
A state-of-the-art 3+1 dimensional cascade + viscous hydro + cascade model vHLLE+UrQMD has been applied to heavy ion collisions in RHIC Beam Energy Scan range $\sqrt\sNN=7.7\dots 200$~GeV. Based on comparison to available experimental data it was estimated that an effective value of shear viscosity over entropy density ratio $\eta/s$ in hydrodynamic stage has to decrease from $\eta/s=0.2$ to $0.08$ as collision energy increases from $\sqrt{\sNN} = 7.7$ to $39$~GeV, and to stay at $\eta/s=0.08$ for $39\le\sqrt{s}\le200$~GeV.

In this work we show how an equation of state with first order phase transition affects the hydrodynamic evolution at those collision energies and changes the results of the model as compared to ``default scenario'' with a crossover type EoS from chiral model.
\end{abstract}

\begin{keyword}
quark-gluon plasma \sep relativistic hydrodynamics \sep hadron cascade

\end{keyword}

\end{frontmatter}


\section{Introduction}
\label{sec-intro}

The goal of ongoing Beam Energy Scan (BES) program at RHIC facility and future experimental programs at GSI FAIR and JINR NICA is to explore the high-$\mu_B$ region of phase diagram of QCD matter and the phase transition from hadron gas to quark-gluon plasma by colliding heavy nuclei at different energies.

Following the success of hydrodynamic description of heavy ion reactions at full RHIC and LHC energies, we have reported on creation of a state-of-the-art viscous hydro+cascade model and its application to heavy ion collisions in the BES collision energy range \cite{Karpenko:2015xea}. In the context of this model,
reproduction of available experimental data requires a finite collision energy dependent shear viscosity over entropy density ratio $\eta/s$ in the hydrodynamic stage. This ratio was found to decrease from $\eta/s=0.2$ to $0.08$ as collision energy increases from $\sqrt{\sNN} = 7.7$ to $39$~GeV, and to stay at $\eta/s=0.08$ for $39\le\sqrt{s}\le200$~GeV. The found collision energy dependence of the effective $\eta/s$ indicates that the physical $\eta/s$-ratio may depend on baryochemical potential, and that $\eta/s$ increases with increasing $\mu_B$.

However, only one version of the equation of state (EoS) was used throughout the analysis in \cite{Karpenko:2015xea}, namely the chiral model EoS \cite{Steinheimer:2010ib}. The question remains: how sensitive the obtained results are to the hydrodynamic EoS and whether it is possible to discriminate between EoSs using experimental data. In this work we show sensitivity of hadronic observables in RHIC BES collision energy range to the choice of the equation of state (EoS) in the fluid stage.

\section{The equations of state in the viscous hydro+cascade model}


Since the model has been described in detail in
Ref.~\cite{Karpenko:2015xea}, we summarise here its main features only.
The initial stage of evolution is described with UrQMD cascade \cite{Bass:1998ca, Bleicher:1999xi}. At a hypersurface of constant Bjorken proper time $\tau=\sqrt{t^2-z^2}=\tau_0$ the system is fluidized, {\it i.e.}~the energy and momentum of individual hadrons are converted into energy and momentum of fluid. The $\tau_0$ is a parameter of the model. At lower RHIC BES energies its value is
set to the time when all initial nucleon-nucleon scatterings have happened. The hydrodynamic stage which follows is modelled with a 3+1 dimensional numerical solution of relativistic viscous hydrodynamics in Israel-Stewart framework using recently developed \texttt{vHLLE} code \cite{Karpenko:2013wva}. Particlization is set to happen when energy density $\epsilon=\epsilon_{\rm sw}$ is reached. The following hadronic stage is described again with UrQMD cascade.

It is worth to note that at BES collision energies local baryon or
charge densities can be large. Therefore a consistent hydrodynamic description requires an EoS which is defined at all physical energy/baryon densities. Presently we have only two such EoSs at hand. One is the chiral model EoS (CM EoS), used in our previous analysis, whereas another is so called ``EoS Q'' \cite{Kolb:2000sd}.

\begin{figure}
\centering
\includegraphics[width=0.52\textwidth]{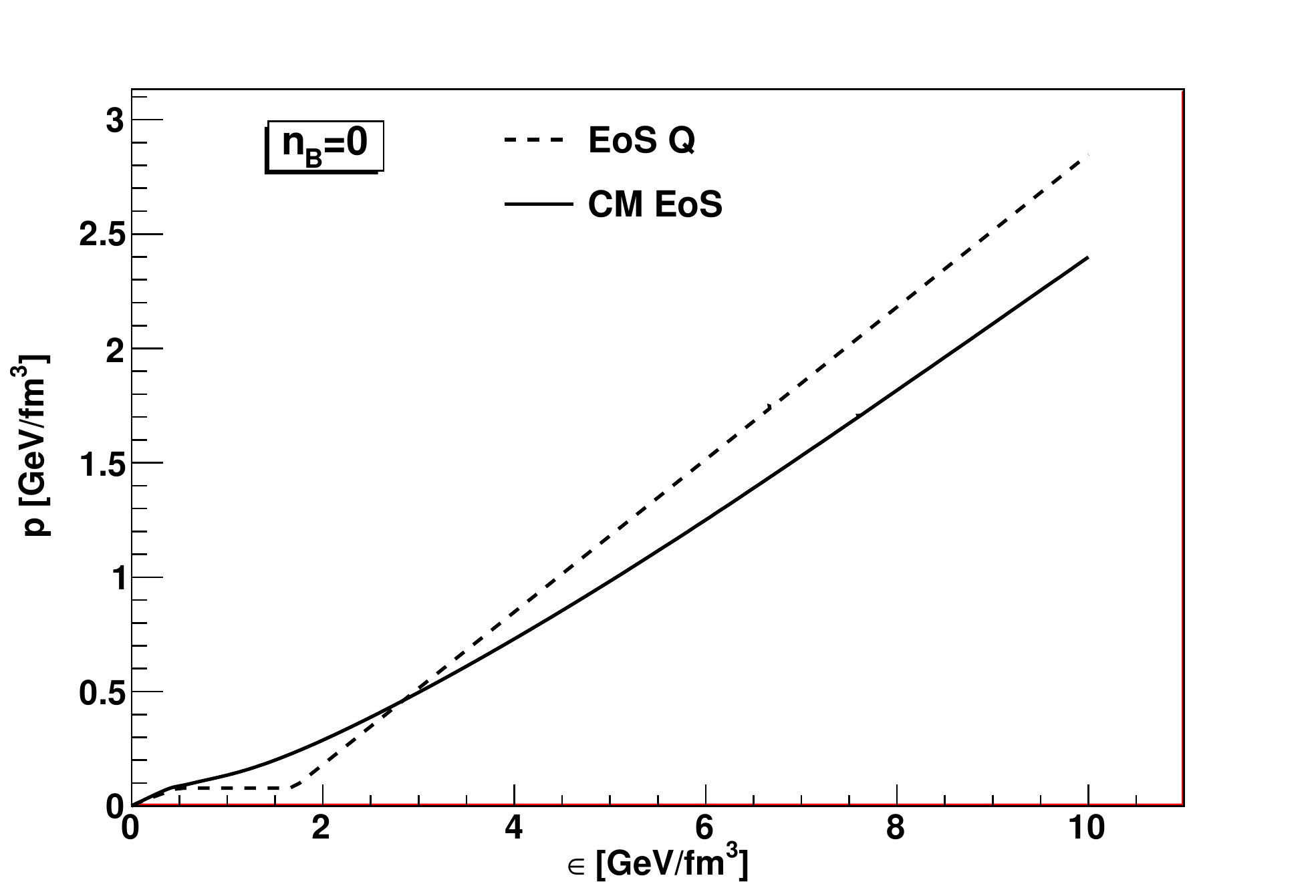}
\caption{Pressure as a function of energy density at zero baryon density, for the two EoSs used in the simulations: chiral model EoS (solid curve) and EoS Q (dashed curve).}\label{fig1}
\end{figure}

\begin{figure}
\includegraphics[width=0.52\textwidth]{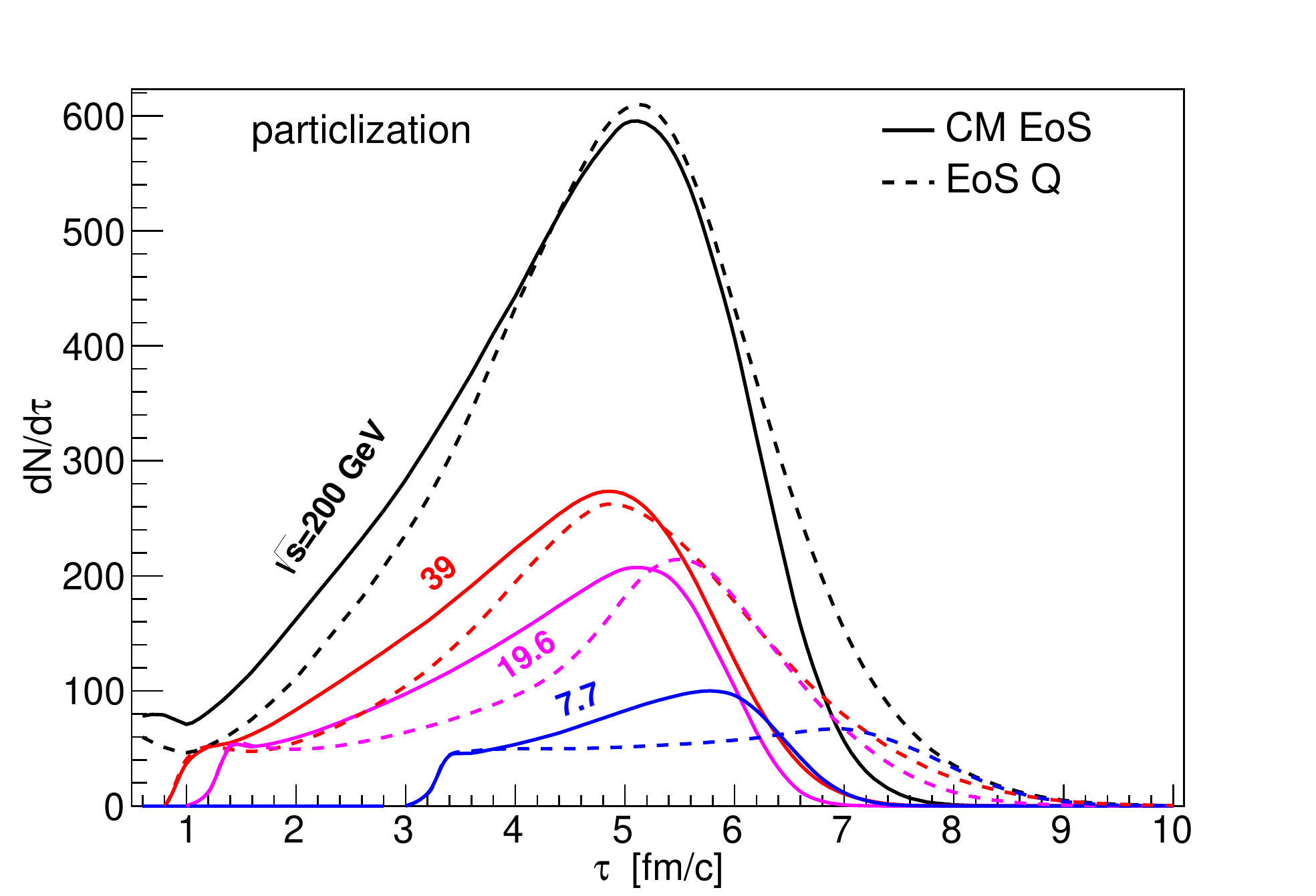}
\includegraphics[width=0.52\textwidth]{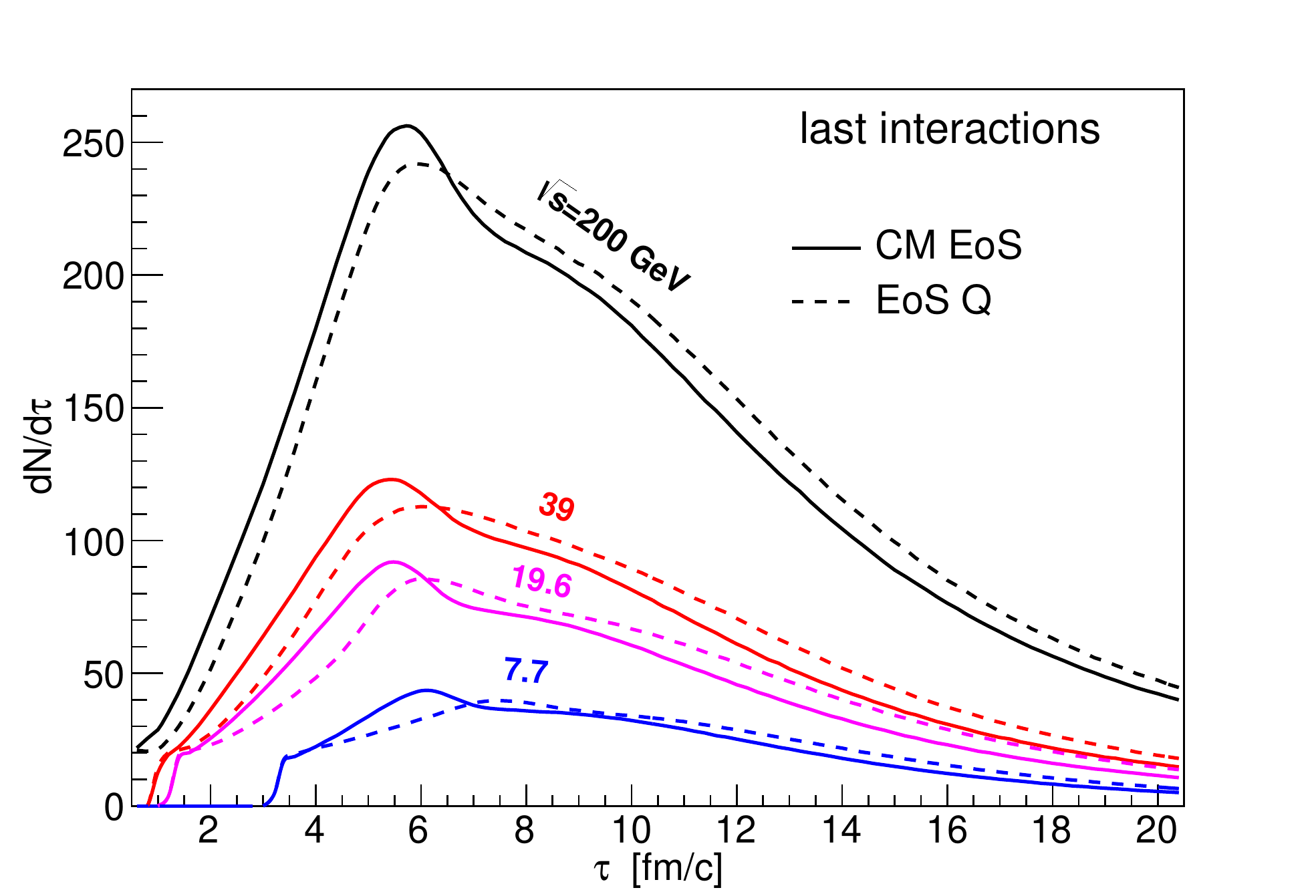}
\caption{Longitudinal proper time distribution of hadrons sampled at particlization surface (left) and their last interaction points (right). All distributions correspond to midrapidity region of 20-30\% central Au-Au collisions at different collision energies, simulated with CM EoS (solid curves) and EoS Q (dashed curves) in the fluid phase.}\label{fig2}
\end{figure}

The chiral model (CM) EoS has correct asymptotic degrees of freedom at the high and low temperature limits -- quarks and hadrons -- and is in qualitative agreement with lattice QCD results at $\mu_B=0$. However, deconfinement transition appears in it as a (wide) crossover where medium modified hadrons coexist with free quarks. The crossover happens at all values of baryon chemical potential.

The EoS Q comprises hadron resonance gas (HG) and quark-gluon plasma (QGP) phases. HG phase is constructed from contributions of hadron resonances with masses up to 2~GeV and includes repulsive interactions via a mean-field potential. QGP phase is described as an ideal gas of massless quarks and gluons inside a large bag with bag constant $B$. The value of the latter is taken to be $B^{1/4}=230$~MeV. The two phases are matched via Maxwell construction, which results in first order phase transition between the phases also at all values of baryon chemical potential. With the given value of the bag constant the transition temperature is $T_c(n_B=0)=164$~MeV at zero baryon density.

The particlization (transition from fluid to individual hadrons) is set to happen at fixed energy density $\epsilon_{\rm sw}=0.5$~GeV/fm$^3$ for both EoS scenarios. At this energy density the system is firmly on the hadronic side for both EoS. We apply conventional Cooper-Frye prescription for hadron distributions on the particlization hypersurface:
\begin{equation}
p^0 \frac{d^3N_i(x)}{d^3p} = d\sigma_\mu p^\mu f(p\cdot u(x),T(x),\mu_i(x)),
 \label{CFp}
\end{equation}
where the phase space distribution function $f$ is taken to correspond to free hadron resonance gas close to local equilibrium with non-equilibrium corrections proportional to the shear stress tensor:
\begin{equation}
f=f_{\rm eq}(p^\nu u_\nu;T,\mu_i)\left[ 1+(1\mp f_{\rm eq})\frac{p_\mu p_\nu \pi^{\mu\nu}}{2T^2(\epsilon+p)} \right].
\end{equation}
Such distribution function is inconsistent with mean fields included in hadron phase of both EoS used in fluid phase. Therefore we recalculate the energy density, pressure, flow velocity $u^\mu$,
temperature, and chemical potentials from the ideal parts of the
energy-momentum tensor and charge currents using a free hadron resonance gas EoS and use these values to
evaluate the particle distributions on the switching surface.

\section{Results and conclusions}

To set up the calculations for the Beam Energy Scan range we take collision energy dependent values of the parameters of the model, used to approach the experimental data with CM EoS. Then we perform two sets of simulations: with CM EoS and EoS Q respectively. First of all we like to see how the hydrodynamic evolution itself is affected by the choice of the EoS. It is difficult to compare individual hydro evolutions with irregular initial conditions and different EoSs. Therefore we visualise the effects of the change of the EoS by using the event-averaged proper time distributions of the hadrons sampled at particlization surfaces, see Fig.~\ref{fig2}, right. From the plot one can find that EoS Q results in longer average duration of the particle emission. This implies that the average duration of the fluid stage (which is defined as a space-time region where $\tau>\tau_0$ and $\epsilon>\epsilon_{\rm sw}=0.5$~GeV/fm$^3$) is longer. Also, the relative change in the average duration of the fluid stage increases with decreasing collision energy and is maximal for lowest collision energy simulated, $\sqrt\sNN=7.7$~GeV. We assume that the prolongation is the effect of the mixed phase in EoS Q, since at higher energy densities it is even harder than CM EoS, see Fig.~\ref{fig1}.

However, after being sampled at the particlization surface, hadrons decay and rescatter in the cascade. From Fig.~\ref{fig2}, left, one can see that the resulting proper time distributions of the last interaction points of hadrons are much wider than the distributions of points of their creation, and differences between the two EoSs in fluid stage are largely smeared. This brings one to the question how much the EoS in the fluid phase affects the final observables.

\begin{figure}
\includegraphics[width=0.52\textwidth]{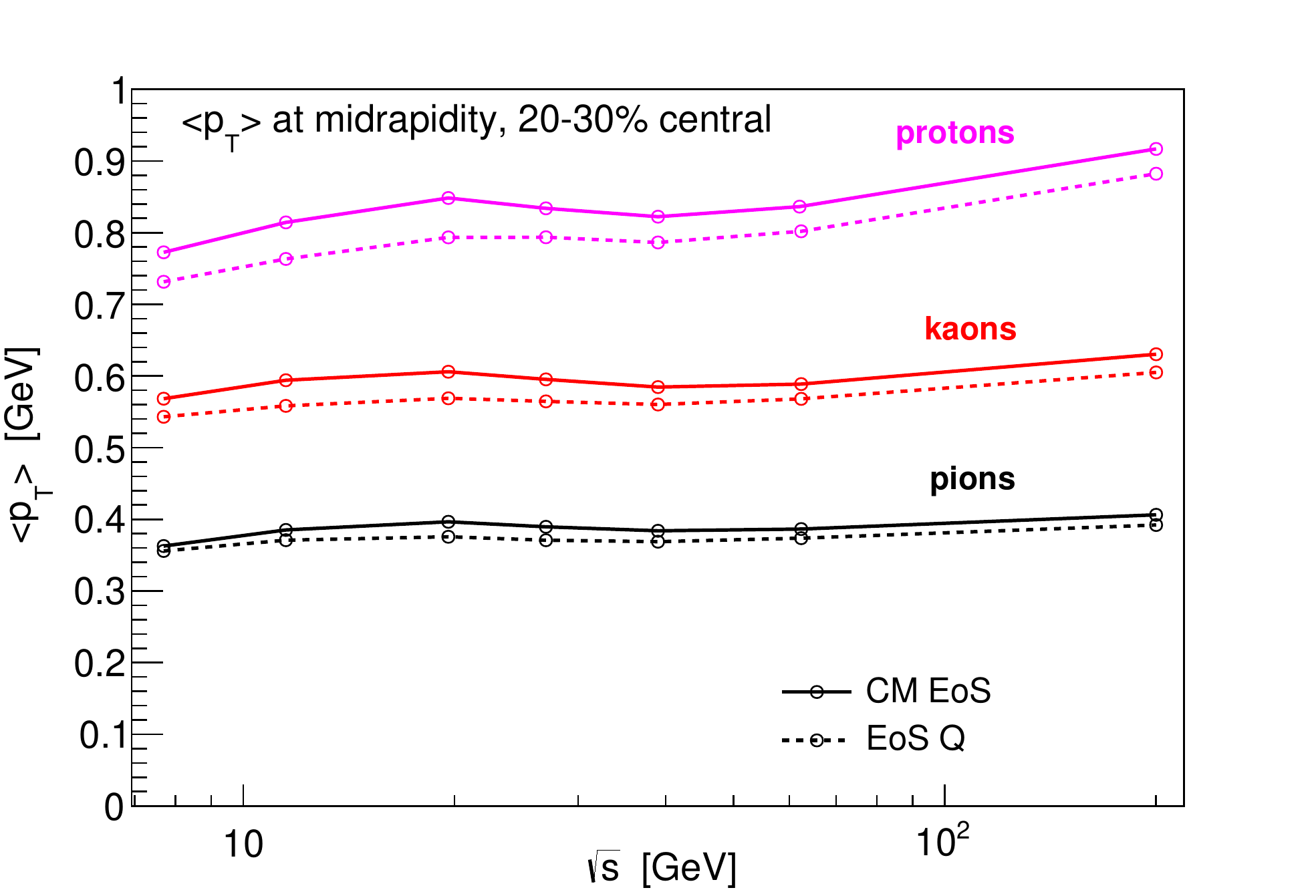}
\includegraphics[width=0.52\textwidth]{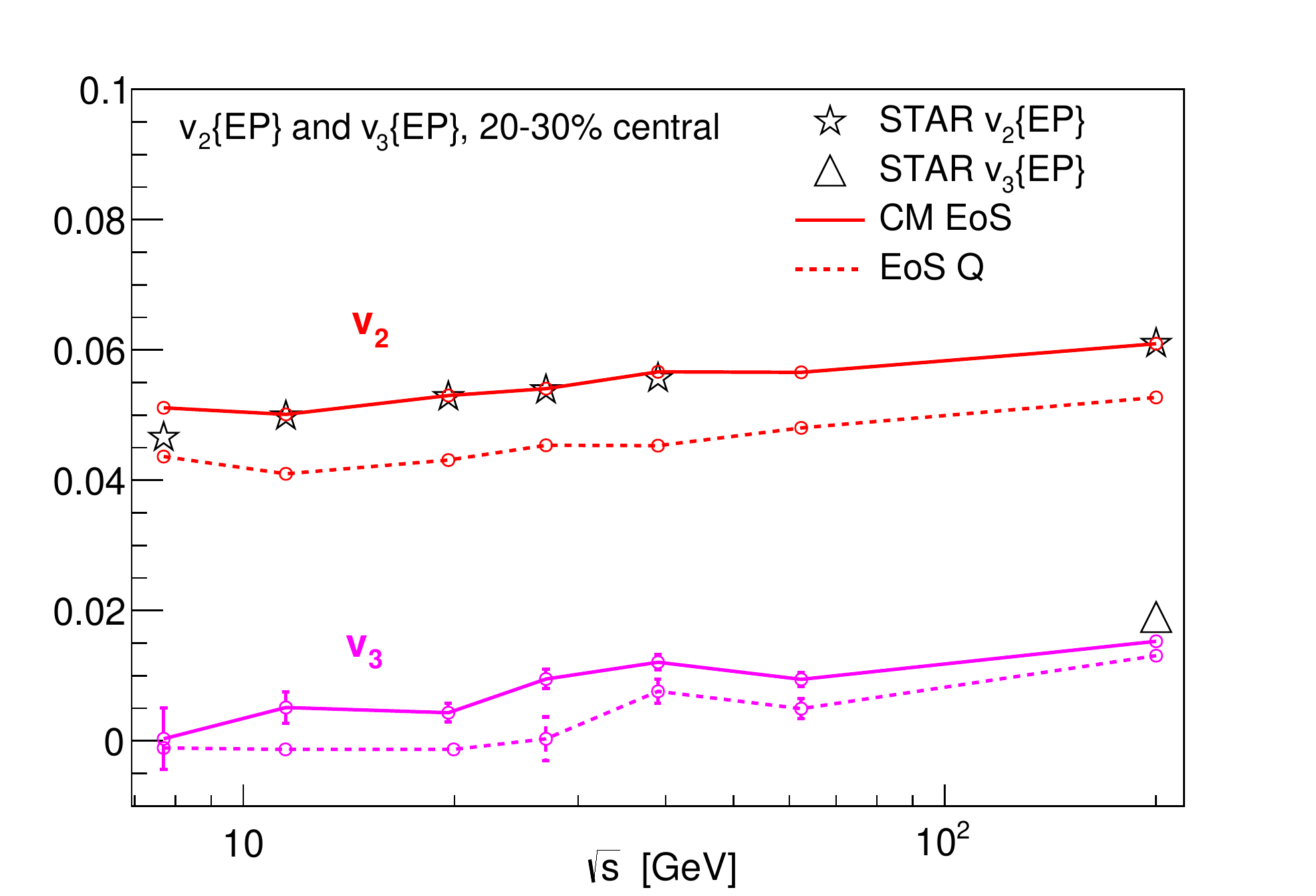}
\caption{Mean $p_T$ of pions, kaons and protons at midrapidity (left), elliptic and triangular flow components of all charged hadrons (right) in 20-30\% central Au-Au collisions, obtained from model simulations with collision energy dependent values of the parameters and two different equations of state in the fluid stage: chiral model EoS (solid curves) and EoS Q (dashed curves). The experimental data points are from the STAR collaboration \cite{Adamczyk:2012ku,Adamczyk:2013waa}.}\label{fig3}
\end{figure}

We found that change in EoS has no impact on the shapes of rapidity distributions of produced particles. There is, however, some impact on the transverse dynamics of the system, which influence transverse momentum distributions of hadrons. From Fig.~\ref{fig3}, left, one can conclude that EoS Q results in some suppression of the average radial flow, which decreases mean $p_T$ of hadrons (larger effect for heavier protons and smaller effect for lighter pions). The largest effect is seen in the $p_T$ integrated elliptic flow, which turns out to be suppressed in EoS Q case by the same amount for all collision energies.

In the previous analysis we have shown that elliptic flow, as well as other observables, can be varied by varying free model parameters. Therefore it remains an open question whether it is possible to readjust the parameters of the model in order to compensate such changes in excitation functions of the elliptic flow and mean transverse momentum while keeping same rapidity distributions.

\section{Acknowledgements}

The simulations have been performed at the Center for Scientific Computing (CSC) at the Goethe-University Frankfurt. The authors acknowledge the financial support by the Helmholtz International Center for FAIR and Hessian LOEWE initiative. HP acknowledges funding by the Helmholtz Young Investigator Group VH-NG-822.  The work of PH was supported by BMBF under contract no. 06FY9092.

\bibliographystyle{elsarticle-num}



\end{document}